\documentclass[fleqn,10pt]{wlscirep}

\usepackage[utf8]{inputenc}
\usepackage[T1]{fontenc}
\usepackage{xcolor}
\usepackage{graphicx}
\usepackage{caption}
\usepackage{subcaption}
\usepackage{libertine}
\usepackage{amssymb}
\usepackage{amsmath,amsthm}
\usepackage{todonotes}
\usepackage{hyperref}

\newcommand{\avg}[1]{\langle{#1}\rangle}
\title{Quantifying the heterogeneous impact of lockdown policies on different socioeconomic classes during the first COVID-19 wave in Colombia}

\author[1,2]{Pablo Valga\~n\'on}
\author[3]{Andr\'es F. Useche}
\author[4,2,*]{David Soriano-Pa\~nos}
\author[5]{Gourab Ghoshal}
\author[1,2,6]{Jes\'us G\'omez-Garde\~nes}
\affil[1]{Departament of Condensed Matter Physics, University of Zaragoza, 50009 Zaragoza (Spain).}
\affil[2]{GOTHAM lab, Institute for Biocomputation and Physics of Complex Systems, University of Zaragoza,
50018 Zaragoza (Spain).}
\affil[3]{Department of Industrial Engineering, School of Engineering, Universidad de Los Andes, 111711 Bogot\'a (Colombia).}
\affil[4]{Instituto Gulbenkian de Ci\^encia, 2780-156 Oeiras (Portugal).}
\affil[5]{Department of Physics and Astronomy, University of Rochester, NY 14627 Rochester, United States).}
\affil[6]{Center for Computational Social Science, University of Kobe, 657-8501 Kobe (Japan).}
\affil[*]{sorianopanos@gmail.com}

\begin{abstract}
In the absence of vaccines, the most widespread reaction to curb COVID-19 pandemic worldwide was the implementation of lockdowns or stay-at-home policies. Despite the reported usefulness of such policies, their efficiency was highly constrained by socioeconomic factors determining their feasibility and their outcome in terms of mobility reduction and the subsequent limitation of social activity. Here we investigate the impact of lockdown policies on the mobility patterns of different socioeconomic classes in the three major cities of Colombia during the first wave of COVID-19 pandemic. In global terms, we find a consistent positive correlation between the reduction in mobility levels and the socioeconomic stratum of the population in the three cities, implying that those with lower incomes were less capable of adopting the aforementioned policies. Our analysis also suggests a strong restructuring of the mobility network of lowest socioeconomic strata during COVID-19 lockdown, which increased their mixing while hampering their connections with wealthiest areas due to a sharp reduction in long-distance trips.
\end{abstract}
\begin{document}

\flushbottom
\maketitle

\thispagestyle{empty}

\section*{Introduction}
 
The lack of approved drugs or vaccines turned non-pharmaceutical interventions~\cite{estrada2020covid,perra2021non,flaxman2020estimating,haug2020ranking} (NPIs) into our best ally to fight COVID-19 pandemic in 2020. NPIs encompass a wide variety of policies implemented to reshape the interaction patterns of the population to hamper the potential contagion pathways of infectious individuals. This goal can be accomplished by targeting different factors involved in the transmission in the virus. At the individual level, the transmissibility of the virus due to contacts between the infected and susceptible population is reduced by promoting prophylactic measures such as the use of masks~\cite{cheng2021face,rader2021mask}. At the population level, the exposure of susceptible individuals to the virus can be reduced by testing-trace-isolate-quarantine (TTIQ) policies~\cite{reyna2021virus,aleta2020modelling,kojaku2021effectiveness}, shortening the effective infectious window of infectious individuals and isolating their contacts, or by population wide lockdowns and stay-at-home policies, decreasing the number of acquaintances of the entire population by controlling their mobility~\cite{arenas2020modeling,di2020impact,marziano2021retrospective}. \\

The success of lockdown or stay-at-home policies in reducing social contacts is not universal but strongly depends on the complex relationship existing between the socioeconomic characteristics of the population and their mobility. In this sense, several studies have found remarkably different mobility patterns across socioeconomic classes classified according to different demographic information such as race, ethnicity or income level~\cite{lotero2016rich,sabatini2006,reme2022quantifying,moro2021mobility,barbosa2021uncovering,hilman2022socioeconomic}. In the context of epidemics, it is worth remarking that, even in uncontrolled scenarios, the coexistence of multiple mobility networks might be highly relevant for the evolution of epidemic outbreaks, leading to heterogeneous epidemic trajectories across the different socioeconomic classes~\cite{soriano2018spreading,bosetti2020heterogeneity}. In addition to the socioeconomic flavor of usual recurrent mobility patterns, the feasibility of the aforementioned policies also varies across different socioeconomic groups. For example, the adoption of stay-at-home policies is clearly related to the possibility of working at home, an option that is much more accessible to those individuals with higher income levels~\cite{castro2022worked,irlacher2021working,bonacini2021working}. Consequently, a positive correlation between the income level of the population and the levels of mobility reduction achieved through lockdown interventions has been measured in different cities~\cite{duenas2021changes,castells2021unequal,aromi2021socioeconomic}.\\

Here we aim at further investigating the interplay between socioeconomic features and the mobility reduction observed as a consequence of lockdown policies in the three major cities of Colombia: Bogot\'a, Medell\'in and Santiago de Cali. As usual in Colombian cities, the population is divided into six different socioeconomic strata according to the quality of their households and neighborhoods. This classification is used to determine the cost of public services and access to government aid. Although a certain degree of overlapping exists between strata and socioeconomic classes, stratum 1 typically gathers those individuals with less economic resources whereas stratum 6 usually correspond to the wealthiest population. By integrating the spatial distribution of the different strata in each city with the time series of mobility flows provided by cell phone data, we first confirm the positive correlation between reduced mobility and the socioeconomic status of the population. We also show that, although the average distance traveled decreased for all strata due to the blocking measures, this effect was much more significant for individuals belonging to the lowest strata as a consequence of the sharp reduction in long-distance trips.\\

Beyond characterizing aggregated statistics for each stratum, here we also want to address how facing a health emergency scenario alter the mixing patterns among residents from different strata.  Our results highlights a heterogeneous microscopic impact of lockdown policies according to the socioeconomic class of the population. Namely, lockdown policies exacerbated both endogenous interactions and segregation of the lowest strata while not having any significant impact on the wealthiest ones.

\section*{Methods}
\subsection*{Construction of the mobility multiplex network}

To estimate the interplay between lockdown policies, socioeconomic information and mobility reduction, we need to construct a multiplex network to capture the different mobility patterns of each stratum in Bogot\'a, Medell\'in and Cali respectively. In particular, these multiplex networks are composed of $L=6$ layers, each one associated with one socioeconomic class (stratum) and $N$ patches, corresponding to the number of areas in which each city is partitioned. In our case these partitions correspond to the spatial resolution of level 12 s2cells, as provided by cell phone data. \\

To construct the mobility network associated with each stratum, we first extract demographic and socioeconomic data from the 2018 Colombian census. As a result, for the three cities here analyzed, we obtain the total number of residents as well as the distribution of households according to their stratum with a resolution of census sector, an administrative unit gathering different census blocks. \\

As for mobility, we extract the mobility patterns of the population from The Google COVID-19 Aggregated Mobility Research Dataset contains anonymized mobility flows aggregated over users who have turned on the Location History setting, which is off by default. This is similar to the data used to show how busy certain types of places are in Google Maps — helping identify when a local business tends to be the most crowded. The data set aggregates flows of people from region to region, which is here further aggregated at the level of 12 s2cells areas, weekly. These correspond to an area between 3.04 and 6.38 squared kilometres. By default, all metrics defined in this section correspond to weekly values, but can be generalized to quantify averages over any arbitrary number of weeks.\\

To produce this data set, machine learning is applied to logs data to automatically segment it into semantic trips \cite{bassolas2019hierarchical}. To provide strong privacy guarantees, all trips were anonymized and aggregated using a differentially private mechanism \cite{wilson2019differentially} to aggregate flows over time (see \href{https://policies.google.com/technologies/anonymization}{https://policies.google.com/technologies/anonymization}). This research is done on the resulting heavily aggregated and differentially private data. No individual user data was ever manually inspected, only heavily aggregated flows of large populations were handled.\\

All anonymized trips are processed in aggregate to extract their origin and destination location and time. For example, if  users traveled from location a to location b within time interval t, the corresponding cell (a,b,t)  in the tensor would be $n\pm err$, where $err$ is Laplacian noise. The automated Laplace mechanism adds random noise drawn from a zero mean Laplace distribution and yields ($\varepsilon, \delta$)-differential privacy guarantee of $\varepsilon=0.66$ and $\delta = 2.1 \times 10-29 $ per metric. Specifically, for each week W and each location pair (A,B), the number of unique users who took a trip from location A to location B during week W is computed. To each of these metrics, Laplace noise is added from a zero-mean distribution of scale 1/0.66. All metrics for which the noisy number of users is lower than 100 are then removed, following the process described in \cite{wilson2019differentially}, and the rest published. This yields that each published metric satisfies ($\varepsilon$,$\delta$)-differential privacy with values defined above. The parameter $\varepsilon$ controls the noise intensity in terms of its variance, while $\delta$ represents the deviation from pure $\varepsilon$-privacy. The closer they are to zero, the stronger the privacy guarantees.\\

To merge socioeconomic information with mobility data, we create $6$ matrices ${\bf f^s}$ ($s=1,...,6$) so that each entry $f^s_{ij}$ encodes the number of individuals from stratum $s$ moving between patches $i$ and $j$. To do this, we must aggregate the information at the census block code level to match the resolution of s2cells so that each urban block is associated in the cell that encapsulates its geometric center. Let us here remark that Colombian cities typically display high levels of segregation among socioeconomic classes~\cite{sabatini2006}, which allows us not to lose detail or aggregate disparate population when merging mobility and socioeconomic data. From census information, the number of residents inside patch $i$ belonging to stratum $s$, $n^s_i$, is computed as:
\begin{equation}
    n_i^s = \sum\limits_{k\in i} E_k^s \tilde{n}_k \;,
\end{equation}
where $\tilde{n}_k$ encodes the number of residents in city block $k$ and $E_k^s$ corresponds to the fraction of households belonging to stratum $s$ inside block $k$. Therefore, the proportion of a certain stratum $s$ in a cell $i$ is calculated as:
\begin{equation}
    R_i^s = \dfrac{n_i^s}{\sum\limits_{s=1}^6{n_i^s}}\;.
\end{equation}\\

To construct the mobility network, we assume that there are no significant differences in the representativeness of each stratum in the mobility data set. This way, we can estimate the mobility network of each stratum (layer) $s$ as: 
\begin{equation}\label{eq:flux}
    f_{ij}^s = f_{ij} R_i^s\ ,
\end{equation}
where $f_{ij}$ represents the origin-destination fluxes estimated from mobility data. The sum of these trips will give us the overall mobility of the population during a week, which reads as follows:
\begin{equation}
    f = \sum_{i,j} f_{ij}\;,
\end{equation}
or, in the case of a particular stratum $s$:
\begin{equation}
    f^s = \sum_{i,j} f_{ij}^s\;.
\end{equation}

\subsection*{Estimating mixing patterns among strata}

Once we have embedded the mobility patterns of the different strata in a multiplex network, we can estimate how mixing between different strata changed over time as a consequence of the implemented lockdown policies. To this aim, we define the entries of the ($6\times 6$) mixing matrix ${\bf M}$, for which each entry $M_{s_1 s_2}$ quantifies the average proportion of individuals belonging to stratum $s_2$ in all destinations visited by individuals from stratum $s_1$. Mathematically we compute these entries as:
\begin{equation}
    M_{s_1 s_2} = \dfrac{\sum\limits_{i,j} f_{ij}^{s_1} R_j^{s_2}}{\sum\limits_{ij} f_{ij}^{s_1}}\label{eq:mixing}\;.
\end{equation}

To compare this metric with the mean-field baseline scenario, in which the population would be well-mixed, we divide these values by the fraction of the population belonging to each stratum. This is a more representative way of analyzing the segregation of the different strata without the bias of the underlying distribution of the population among them. The entries of the re-normalized mixing matrix, ${\bf M_N}$, are thus defined as:
\begin{equation}\label{eq:matrix_norm}
    {M_N}_{s_1 s_2} = \dfrac{M_{s_1 s_2}}{n^{s_2}}
\end{equation}
where $n^{s_2}$ is the total population that belongs to stratum $s_2$ such that
\begin{equation}
    n^s = \frac{\sum\limits_{i}n_i^{s}}{\sum\limits_{i,s'} n_i^{s'}}\;.
\end{equation}

Each row of the mixing matrix provides information about the segregation of the stratum $s_1$, which we can quantify in a single metric using Gibbs entropy formula. This value will grow larger the less segregated the stratum is, as the components of the row become more homogeneous. It is given by the equation
\begin{equation}
    S_{Gibbs}(s_1) = - k_B \sum_{s_2} M_{s_1 s_2} \ln{(M_{s_1 s_2})}\;.
\end{equation}

Another metric that will provide us with useful information about how people from different strata mix is the average distance that any person needs to travel to reach her destination.  This distance can change over time and depends on the stratum of the individual. In particular, the average distance an individual from stratum $s$ travels, denoted in what follows by $\avg{d^s}$ is calculated as:
\begin{equation}
    \avg{d^s} = \dfrac{\sum_{ij} d_{ij} f_{ij}^s}{\sum_{ij}f_{ij}^s}\,,
\end{equation}
where $d_{ij}$ is the geometric distance between the centroids of geographical areas $i$ and $j$. Analogously, the distance a person from stratum $s_1$ travels on average to arrive at a stratum $s_2$ destination, hereinafter denoted by $D^{s_1s_2}$, is computed as
\begin{equation}\label{eq:strata_distances}
    D^{s_1 s_2} = \dfrac{\sum\limits_{i,j} d_{ij}f_{ij}^{s_1}R_j^{s_2}}{\sum\limits_{i,j} f_{ij}^{s_1} R_j^{s_2} }\;.
\end{equation}

\subsection*{Estimation of the effective reproductive number $R_t$}
The estimation of the instantaneous basic reproductive number $R_t$ is obtained using the EpiEstim package in R \cite{epiEstim}, assuming, as estimated by Ganyani et al. \cite{ganyani2020estimating}, that the generation time distribution follows a gamma function with mean 5.2 (CI: 3.78 - 6.78) days and standard deviation 1.72 (CI: 0.91 - 3.93) days. We assume a standard deviation for the mean and the standard deviation of the generation time of 1.0 days and 1.2 days, respectively. In addition, we consider a rolling average of 7 days for the estimation, whereas the intervals are calculated by averaging over 100 samples of generation time distributions and considering 100 posteriors for each one of them.\\

\begin{figure}[t!]
    \centering

    \includegraphics[width=.85\textwidth]{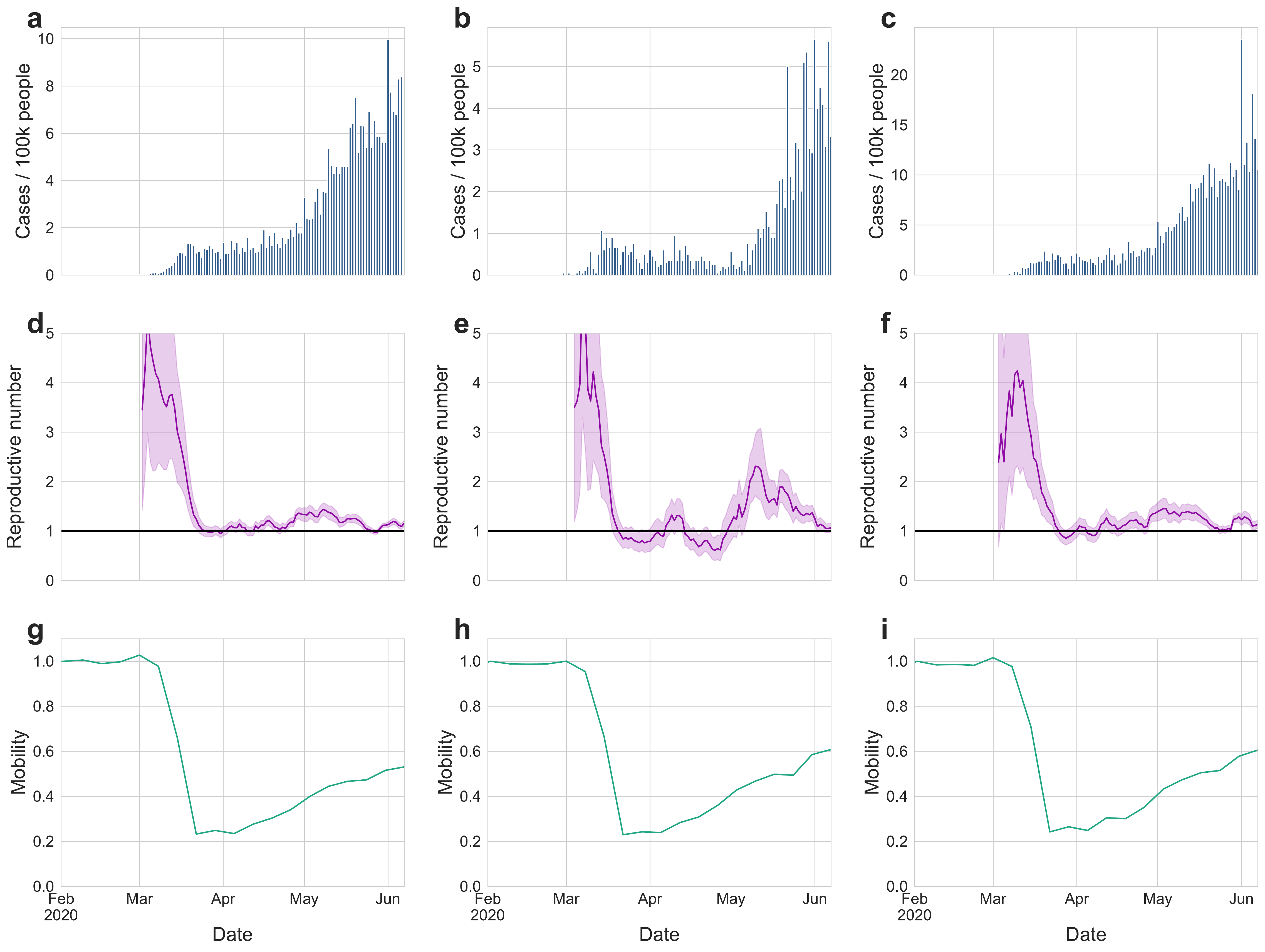}

    \caption{\textbf{a}-\textbf{c} Time evolution of the reported cases per 100k inhabitants  according to the symptoms onset date. \textbf{d}-\textbf{f} Time evolution of the effective reproductive number $R_t$. \textbf{g}-\textbf{i} Time evolution of the aggregated number of trips across the city. For the sake of comparison, the number of trips is re-scaled by those corresponding to a pre-pandemic scenario, i.e. those made during the week starting from 2020-02-02. From left to right, the information shown in all rows corresponds to Bogot\'a, Medell\'in and Santiago de Cali respectively.}
    \label{fig:reproductive_number}
\end{figure}

\section*{Results}
\subsection*{NPIs and epidemic curves in Colombia}
We first aim at addressing whether NPIs promoted by local authorities were helpful to contain the growth of COVID-19 cases in Colombia. The first confirmed case in Colombia was notified on the 6th of March of 2020. By that time, the government had already recommended the residents to work at home, but the first mandatory confinement was not put in place until the 17th of March, when a state of emergency was declared focused on protecting vulnerable groups of people. The former state of emergency was strengthened on March 24, where the first nationwide lockdown started. \\

\begin{figure}[t!]
\centering
\includegraphics[width=.87\textwidth]{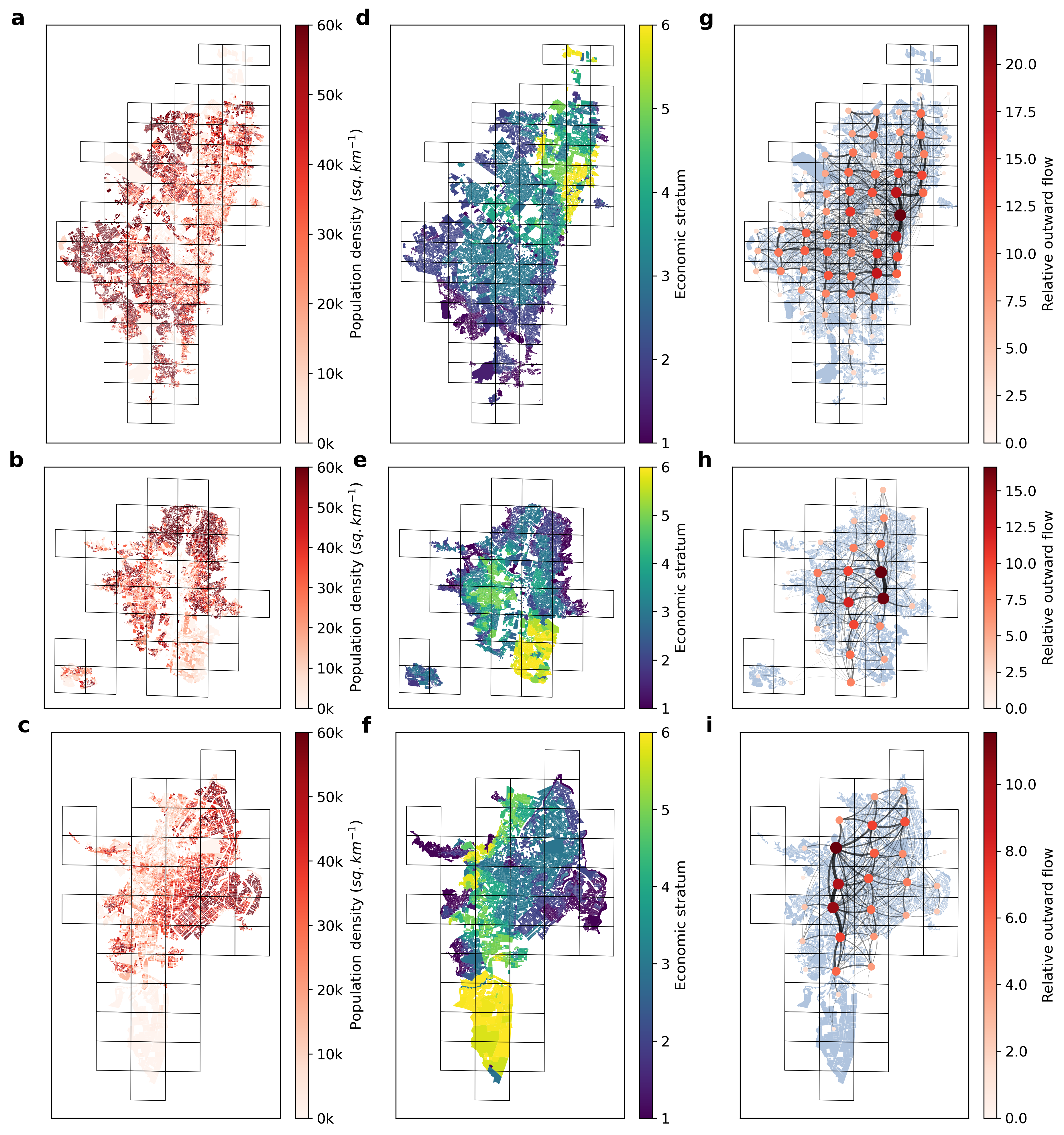}
\caption{\textbf{a}-\textbf{c} Population density maps at the level of census sector, corresponding to the 2018 census. \textbf{d}-\textbf{f} Average economic strata of the different households at the level of census sectors, ranging from stratum 1 typically gathering those individuals with lower economic income to stratum 6 associated with wealthiest population. \textbf{g}-\textbf{i} Schematic representation of the mobility network of each city with a spatial resolution of level 12 s2cells (see Methods for further explanations). Both color and size of nodes are proportional to the total number of trips departing from a given area whereas the edge thickness reflects the the number of trips recorded between two locations. From top to bottom, the information shown corresponds to Bogot\'a, Medell\'in and Santiago de Cali respectively.}
\label{fig:resolucion}
\end{figure}

Figure~\ref{fig:reproductive_number} illustrates how these different interventions shaped the course of COVID-19 epidemic trajectory in the three cities here analyzed. In particular, it shows how the policies implemented from mid March onwards led to a decrease of the speed at which COVID-cases grew in these cities (Figs.~\ref{fig:reproductive_number}.a-c) reflected by a sharp reduction of the effective reproductive number $R_t$ (Figs.~\ref{fig:reproductive_number}.d-f), which quantifies the number of secondary infections made by an average infectious individuals at a given day $t$. We estimate the evolution of such quantity by using the {\it EpiEstim} package in R, as detailed in the Methods section. The effectiveness of lockdown policies is also reflected in the reduction of population movements with respect to the baseline mobility computed before the COVID-19 pandemic, as shown in Figs.~\ref{fig:reproductive_number}.g-i. Although the qualitative trends are very similar among the three cities, let us note how the earlier lockdown policies promoted by local authorities in Medell\'in reduced significantly the burden caused by the pandemic in its initial stage. This highlights the importance of an early reaction to the exponential growth of cases at the beginning of an outbreak, as pointed out by different studies analyzing the impact of NPIs during the COVID-19 pandemic~\cite{marziano2021retrospective,di2020impact,arenas2020modeling,steinegger2021retrospective}.  \\

\begin{figure}[!h]
    \centering
    \includegraphics[width=.95\textwidth]{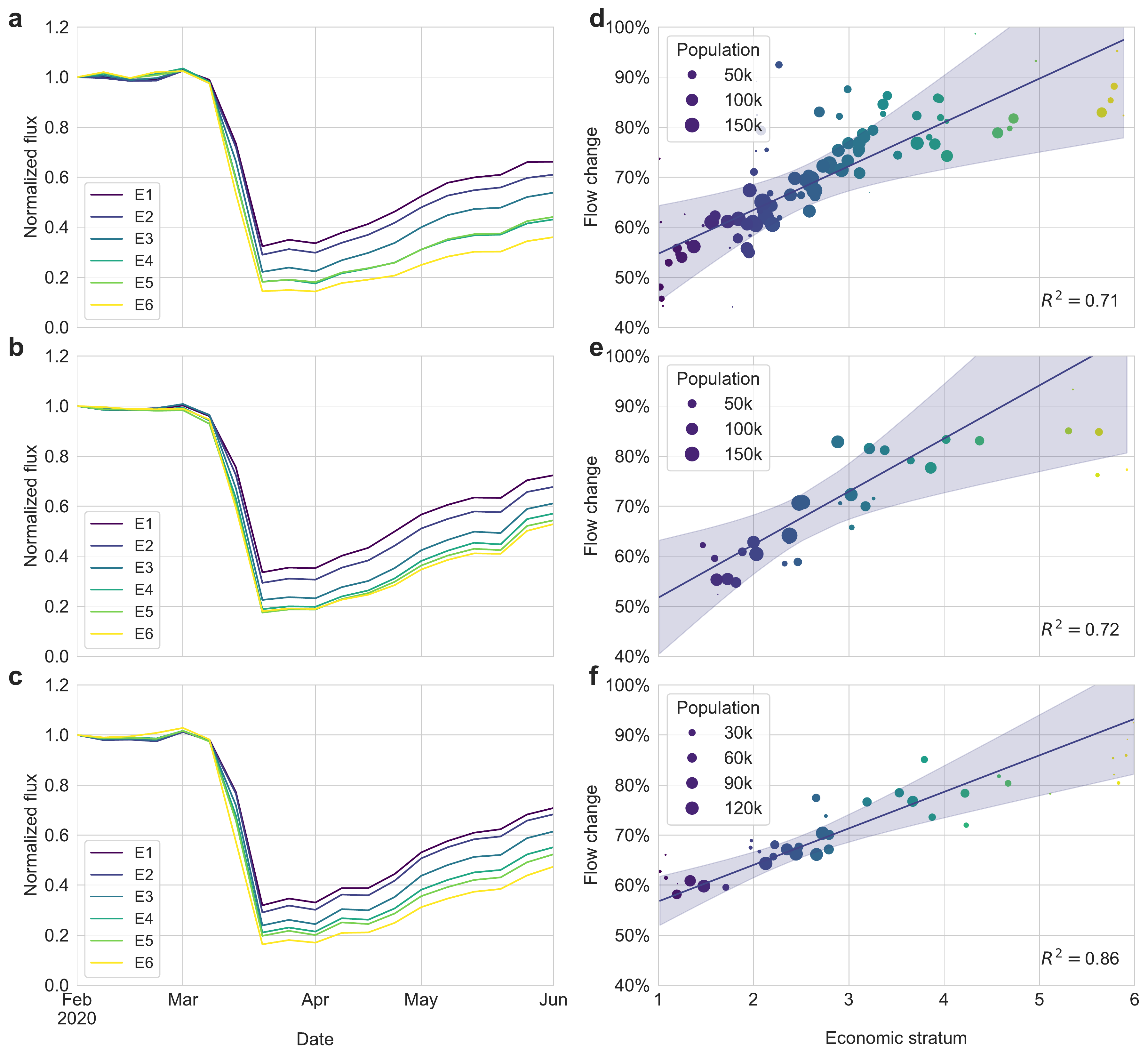}
    \caption{\textbf{a}-\textbf{c} Time evolution of the aggregated number of trips for each stratum (color code), re-scaled to a reference value set on 2020-02-02, corresponding to the pre-pandemic scenario. \textbf{d}-\textbf{f} Reduction in the trips departing from each patch on the week starting from 2020-03-29 compared to a baseline scenario (2020-02-02) according to the average economic stratum of its residents. The size of the dots denotes the number of residents in the corresponding patch whereas the color encodes the stratum information ranging from poorest areas (blue) to the richest ones (yellow). The shaded area is the prediction interval of the linear regression, and it represents the standard error of the predictions from the model, obtained with the member \emph{var\_pred\_mean} from the Python library \emph{statsmodels}. From top to bottom, the information shown corresponds to Bogot\'a, Medell\'in and Santiago de Cali respectively.}
    \label{fig:estratos_tiempo}
\end{figure}

\subsection*{Mobility, lockdown and socioeconomic classes}

Now we move to analyze the socioeconomic determinants shaping the relationship between lockdown policies and mobility reduction which has been discussed from a global perspective in the former section. To do so, we construct the mobility network associated with each stratum by merging the mobility patterns with socioeconomic information provided by census surveys as detailed in the Methods section. For the sake of illustration, we represent in Fig.~\ref{fig:resolucion} the spatial distribution of: the number of residents (panel a), the household stratum (panel b), and the aggregated mobility patterns (panel c) for each of the three cities here analyzed.\\

\begin{figure}[t!]
\centering
    \includegraphics[width=0.87\textwidth]{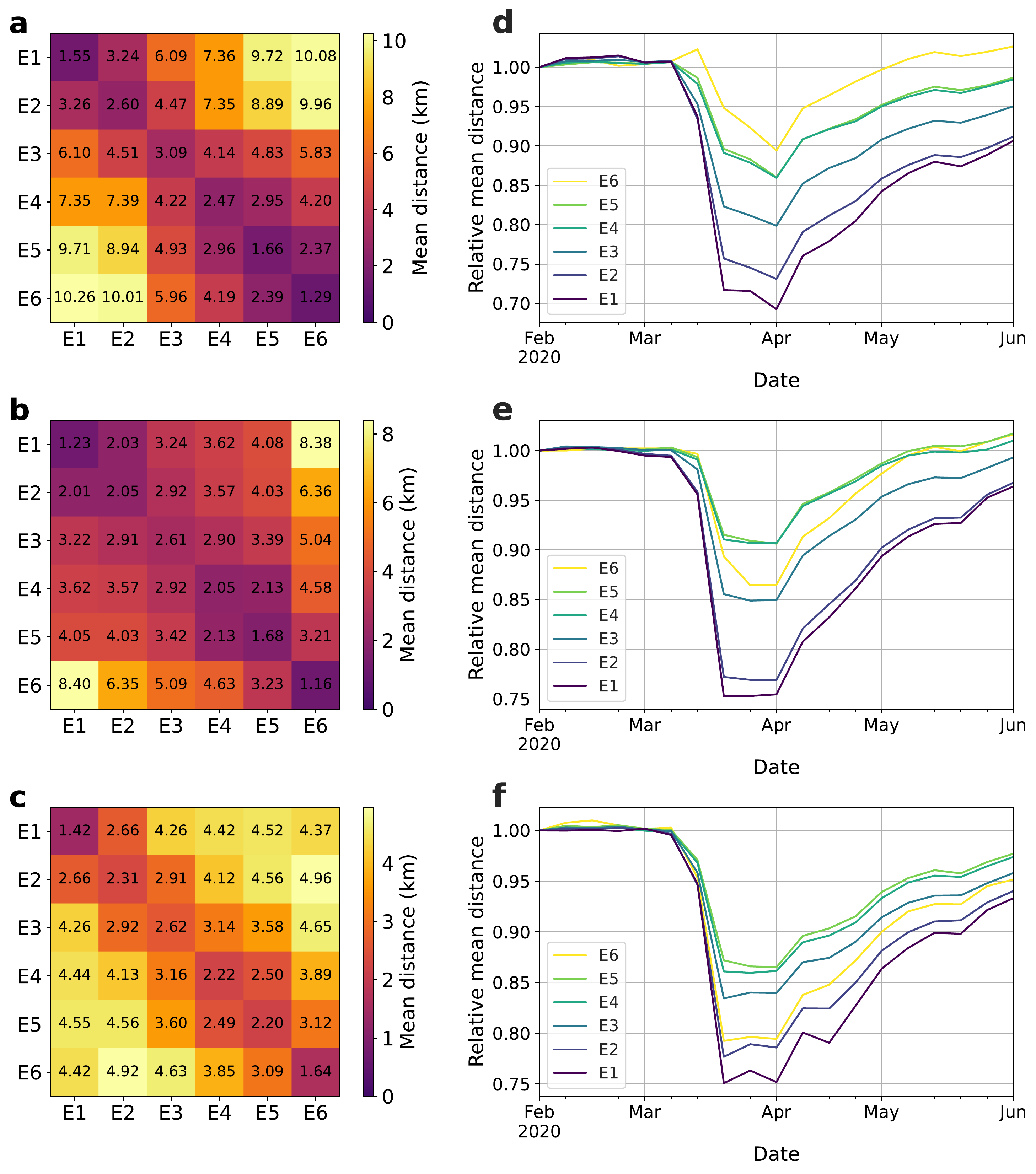}
    \caption{\textbf{a}-\textbf{c} Distance matrices ${\bf D}$ encoding the average distance of trips connecting areas corresponding with two given strata. \textbf{d}-\textbf{f} Time evolution of the average distance of travels made by individuals belonging to each stratum $s$, $\avg{d^s}$ (color code). Note that the values represented have been re-scaled by those corresponding to the pre-pandemic scenario. From top to bottom, the information shown corresponds to Bogot\'a, Medell\'in and Cali respectively.}
    \label{fig:mean_median}
\end{figure}

Once constructed the multiplex mobility network, we can repeat the analysis shown in Figs.~\ref{fig:reproductive_number}.g-h by focusing on the reduction in the level of mobility of each stratum. Figures~\ref{fig:estratos_tiempo}.a-c clearly show how mobility reduction was not homogeneous across strata, being much more significant for the higher strata typically gathering socioeconomic classes with higher income. To further quantify this phenomenon, we compute the reduction in the volume of flows due to lockdown policies by comparing mobility data corresponding to the first week of April and the first of February recorded for each geographical area (patch). Figures~\ref{fig:estratos_tiempo} d-f confirm the positive correlation between the average economic stratum in each patch and the reduction in mobility achieved by its residents. This correlation is consistent among the three cities here analyzed, being the Spearman correlation coefficients between both variables $\rho_S=0.832$, $\rho_S=0.878$ and $\rho_S=0.924$ for Bogotá, Medell\'in and Santiago de Cali respectively.\\

\subsection*{Lockdown restructures mobility networks}

So far, we have tackled the connection between lockdown policies and mobility by studying how the macroscopic volume of movements associated with each stratum varied as these policies were enforced. In this section, we are interested in addressing whether, beyond the overall reduction in the level of mobility across the city, these interventions modified substantially the microscopic structure of the mobility network of each stratum. To do so, we focus on two different indicators: the distance involved in each stratum's trips and the socioeconomic composition of their destinations. \\

\begin{figure}[t!]
\centering
\includegraphics[width=1\textwidth]{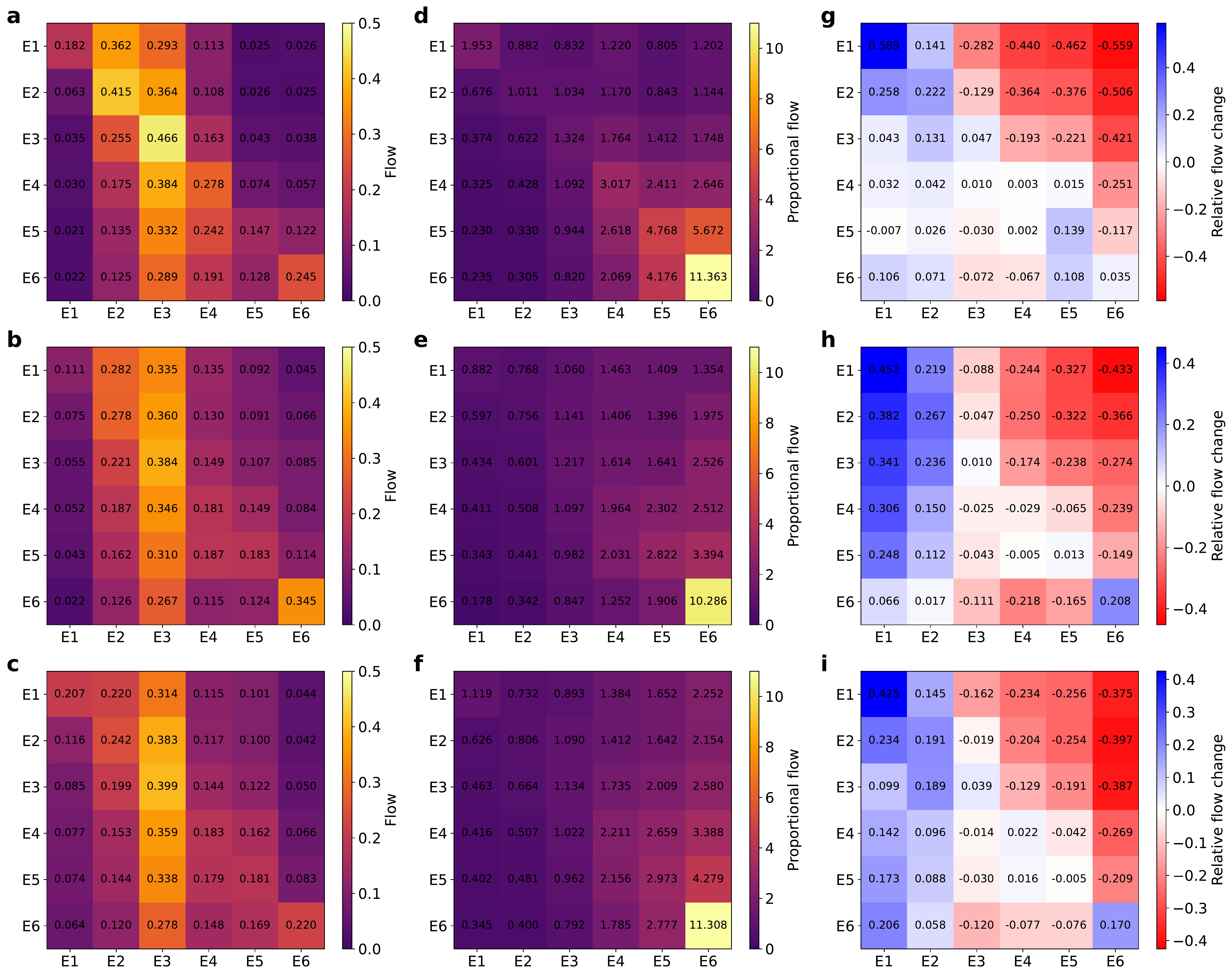}
\caption{\textbf{a}-\textbf{c} Mixing matrices ${\bf M}$ encoding the fraction of trips corresponding to a certain stratum (rows) arriving in areas associated with another stratum (columns) on 2020-02-02 (pre-pandemic scenario). The sum of the elements on each row is equal to one. \textbf{d}-\textbf{f} Re-normalized mixing matrices ${\bf M_N}$ based on the proportion of the population belonging to each stratum, as calculated in equation \eqref{eq:matrix_norm}. In these matrices an entry with value larger than one implies that there is a larger tendency to contact that stratum than what expected in a mean-field scenario. \textbf{g}-\textbf{i} Relative change of these matrices between 2020-02-02 and  2020-03-29 (lockdown scenario). From top to bottom, the information shown corresponds to the cities of Bogot\'a, Medell\'in and Santiago de Cali respectively.}
\label{fig:matrices}
\end{figure}

Before studying the impact of lockdown policies, let us characterize the structure of the mobility networks in the baseline scenario (February 2020) before the arrival of COVID-19 pandemic in the three cities here analyzed. For this purpose, we compute the matrix ${\bf D}$, where each element $D^{s_1s_2}$ of this matrix is the average distance of the flows from the zones associated with stratum $s_1$ to those associated with stratum $s_2$ (see Methods for the computation of the matrix). Figures.~\ref{fig:mean_median}.a-c show that the different strata are systematically segregated in the cities studied, since the distance between two strata is greater the further apart they are from a socioeconomic point of view. These results, computed from the merged mobility networks, are consistent with the clear segregation observed in Figs.\ref{fig:resolucion}.d-f showing the spatial distribution of strata across these cities with the finest spatial resolution of census sectors.\\

To compute the impact of lockdown policies, we represent in Figs.~\ref{fig:mean_median}.d-f the time evolution of the average distance of trips departing from areas associated with each stratum $s_1$, denoted by $\avg{d^{s_1}}$ (see Methods for the derivation of this quantity). In these panels we observe a robust trend across the three cities consisting in a drastic reduction of trip distance as soon as NPIs were implemented. Note, however, that this effect is not homogeneous across each city, being more pronounced for those trips departing from areas associated with lower strata. \\

To unravel the roots of the heterogeneous trends observed in Figs.~\ref{fig:mean_median}.d-f, we investigate how lockdown policies altered the socioeconomic composition of the destinations visited by each stratum. This information is encoded in the mixing matrix ${\bf M}$ in which each element $M_{s_1 s_2}$ denotes the average fraction of individuals belonging to stratum $s_2$ in the destinations of the trips departing from stratum $s_1$ areas. The mathematical expression of these elements can be found in the Methods sections. As for the distance analysis, let us first analyze the mixing matrix ${\bf M}$ in the baseline scenario, when no lockdown policies were at play. Figures~\ref{fig:matrices}.a-c confirm the strong socioeconomic impact on mixing patterns in Colombian cities. Specifically, we observe that trips connecting areas populated by similar strata are overrepresented whereas connections among distant strata are hampered. 
By removing the bias related to the unequal population size of the different strata and computing the re-normalized mixing matrix ${\bf M_N}$ (see Methods for details on its calculation), the former result becomes more evident, as shown in Figs~\ref{fig:matrices}.d-f.\\

\begin{figure}[t!]
\centering
\includegraphics[width=1\textwidth]{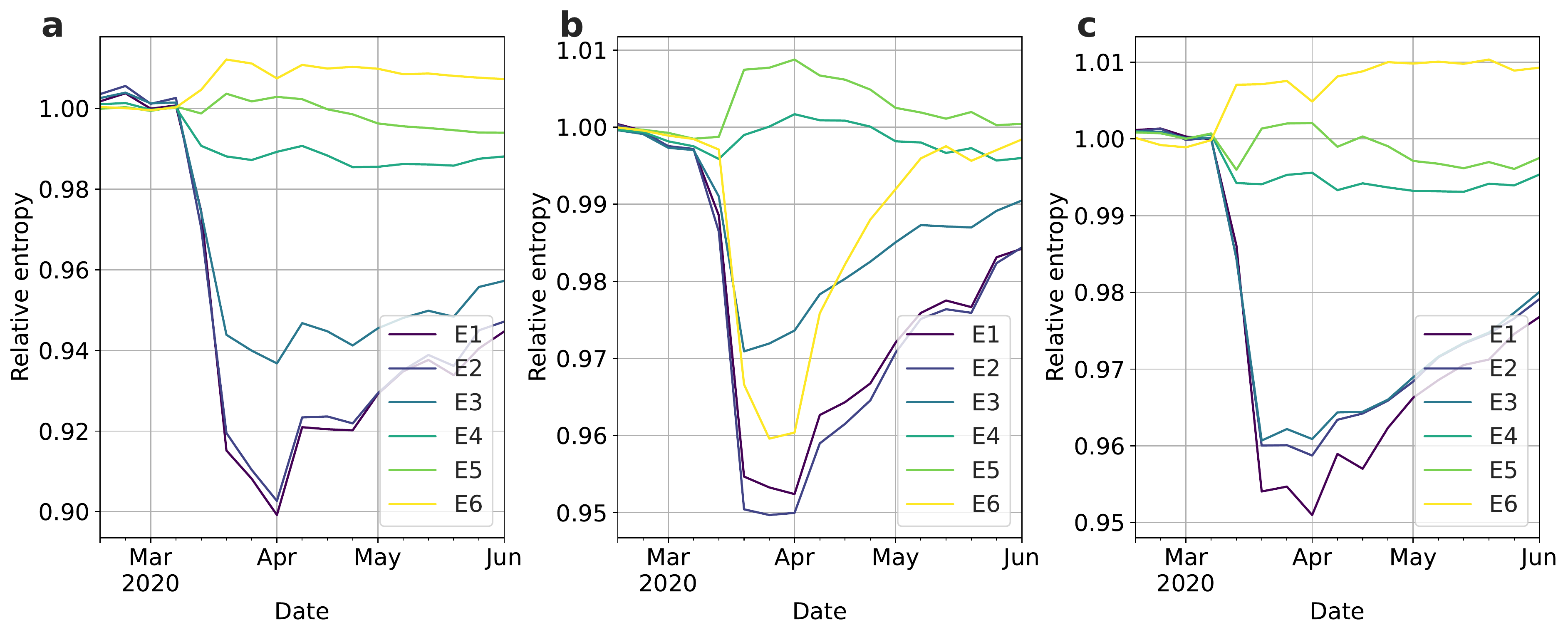}
\caption{Time evolution of the Gibbs entropy for each row of the mixing matrix, which encodes the socioeconomic structure of the destinations visited by each stratum (color code), in Bogot\'a ({\bf a}), Medell\'in ({\bf b}) and Cali ({\bf c}) respectively. Let us note that the lower the value of the Gibss entropy, the more concentrated the trips are towards a small number of economic strata.}
\label{fig:entropy}
\end{figure}

In Figs.~\ref{fig:matrices}.g-i we show how lockdown policies altered the elements of the mixing matrices. There, by comparing the structure of the matrices corresponding to the baseline scenario and the week starting from 29th March (during lockdown), we observe important heterogeneities on the impact of the implemented policies on the mobility patterns across different strata. Once again, these heterogeneities across strata are consistent across the three cities here studied. In the first place, confinement policies provoked important modifications in the mobility patterns of lowest strata, exacerbating their endogenous nature while making it difficult for them to mix with those belonging to higher strata. Taking this result together with the average trip distance connecting different strata shown in Fig. \ref{fig:mean_median}, we realize that the pronounced decrease in the trip distance for the lowest strata as a result of the blocking policies is explained by the huge reduction in long-distance trips connecting them to areas populated by the higher strata. In contrast, lockdown policies seemed to have a less relevant impact on the structure of contacts of those with more economic resources in Colombia. Note however that, despite this overall small variation, lockdown policies appeared to increase the diversification of mobility patterns of those belonging to higher strata, which moved more evenly to places belonging to socially distant strata.\\

To round off our study, we compute the time evolution of the Gibbs entropy of the entries of the mixing matrices corresponding to each stratum and represent them in Fig.~\ref{fig:entropy}. The entropy values shown in these figures are normalized by their values as of 02-02-2020 (before the NPIs came into force). The three panels confirm the previous results: lockdown policies reduced considerably the social variability of the destinations visited by lower strata whereas it left almost unaltered those corresponding to higher ones. However we find an exception in Medell\'in, where population from stratum 6 tend to isolate more as also shown by Fig.~\ref{fig:matrices}h.\\

\section*{Discussion}

The extremely rapid unfolding of COVID-19 pandemic at the beginning of 2020 arose from the subsequent combination of the international exportation of infectious cases~\cite{gilbert2020preparedness,chinazzi2020effect} through the airport mobility network and the local exponential growth in the number of cases driven by community transmission of the virus. Since vaccines were not available at the time, many countries deployed NPIs to reduce the impact of the pandemic. These containment measures included mobility restrictions to curb the importation of cases from high-incidence areas and local control policies that reshaped the contact patterns of the population responsible for community transmission of the virus.\\

In this work, we have studied how the NPIs altered the mobility patterns of different socioeconomic classes in the three main cities of Colombia. Our analysis reveals a strongly unequal impact of control policies on the different strata into which Colombian society is divided. First, we have reported a consistent positive correlation between the reduction in the number of trips registered for a given area and the average stratum of its residents, implying that those individuals with more economic resources, typically belonging to higher strata, were more able to reduce their mobility and, therefore, their social activity. Secondly, we have revealed that lockdown policies, in addition to affecting the volume of movements recorded in each city, modified their architecture, reducing the distance involved in the movements and increasing the social similarity in the composition of the origin and destination of the flows recorded. However, this phenomenon was not homogeneous in all strata. The blocking policies hindered long-distance urban trips connecting lower socioeconomic classes with wealthier areas, leading to a significant increase in their endogenous interactions with geographically proximate individuals and a reduction in the usual distance of their trips. In contrast, our analysis reveals that lock-in policies hardly affected the mobility patterns of the higher strata. \\

Our results suggest the need for accounting for socioeconomic variables when assessing the feasibility or the expected impact of control policies to mitigate an epidemic outbreak. In this sense, despite the clear improvement of the health situation while gaining time for vaccine development, the socioeconomic dimension of NPIs entailed undesired collateral effects such as economic crisis~\cite{maital2020global,albanesi2021effects} or the increase of social inequalities in different countries as a result of lockdown policies~\cite{van2020covid,wachtler2020socioeconomic,bajos2021lockdown}. Beyond social implications, the exacerbation of social inequalities might also have important implications for the evolution of epidemic outbreaks, as proven by the influence of income gradients on COVID-19 associated mortality~\cite{arceo2022income,drefahl2020population,decoster2021income}. We hope that our study sheds light into the interplay between socioeconomic information and epidemic spreading and paves the way to the design of control policies optimizing the trade-off between their health outcome and the damage of the socioeconomic fabric derived from their implementation.\\

\subsection*{Limitations}
Our results should be interpreted in light of several limitations. First, the Google mobility data is limited to smartphone users who have opted in to Google’s Location History feature, which is off by default. Importantly, these limited data are only viewed through the lens of differential privacy algorithms, specifically designed to protect user anonymity and obscure fine detail. Furthermore, comparisons across rather than within locations are only descriptive since these regions can differ in substantial ways.\\

Another limitation of this data set is that the origin of one movement recorded for one individual does not necessarily coincide with its residence, which might introduce some noise when computing the reduction of the mobility associated to each stratum. Nonetheless, previous studies leveraging this kind of data sets have found comparable results to those obtained when analyzing mobility patterns coming from census surveys~\cite{barbosa2021uncovering,bassolas2019hierarchical}. \\

Moreover, to construct the multiplex mobility networks, we have assumed that all socioeconomic strata are proportionally represented in the data set, thus neglecting socioeconomic biases in the use of mobile phone devices, which may vary by location. While this assumption could remain controversial, our results are consistent with those obtained by analyzing variations in public transport usage as a result of lockdown policies~\cite{duenas2021changes}. \\

Finally, we have restricted our analysis to cities in Colombia, where socioeconomic classes are highly segregated. This segregation is partly explained by the spatial distribution of these populations among cities, where poorer populations tend to be located on informal settlements along the cities' boundaries. Frequently, these far locations tend to be more disconnected from city centers and economical hubs due to scarce access to transport services. The generality of our findings when extending our analysis to other societies deserves further investigation and remains as future work.

\section*{Acknowledgements}

J.G.-G. and P.V. acknowledge funding from grant PID2020-113582GB-I00 funded by MCIN/AEI/10.13039/501100011033; and by the Departamento de Industria e Innovaci\'on del Gobierno de Arag\'on y Fondo Social Europeo through Grant No. E36-20R (FENOL group).

\section*{Author contributions statement}

DSP, GG and JGG conceived the study. PV and AU analyzed the data. All the authors discussed the results and wrote the manuscript.

\section*{Data availability}
Mobility data are extracted from the Google SARS-CoV-2 Aggregated Mobility Research Dataset and are available with permission from Google LLC. Data regarding the demographic and socioeconomic information are publicly available and provided by the national statistics department in Colombia (https://www.dane.gov.co/index.php/estadisticas-por-tema/demografia-y-poblacion/censo-nacional-de-poblacion-y-vivenda-2018). Data about the time evolution of COVID-19 cases are extracted from official data reported by the authorities in Colombia (https://www.datos.gov.co/Salud-y-Protecci-n-Social/Casos-positivos-de-COVID-19-en-Colombia/gt2j-8ykr/data). The datasets used and/or analysed during the current study are available from the corresponding author on reasonable request.

\section*{Code availability}
Codes used to generate the figures of the manuscript are available from the corresponding author on reasonable request.

\section*{Additional information}
The authors declare no competing interests.

\end{document}